\documentclass[conference,10pt]{IEEEtran}
\normalsize

\usepackage{graphicx}
\usepackage{epstopdf}

\usepackage{comment}
\usepackage{bbm} 
\usepackage{amsfonts} 
\usepackage{amssymb}  % for \triangleq  and \blacksquare
\usepackage{amsmath} % for \boldsymbol
\usepackage{enumitem}
\usepackage{multicol}
\usepackage{mathtools}
\usepackage{cuted}

\newcommand{\SMR}{\mbox{SINR}^{\sst{MR}}}   
\newcommand{\SZF}{\mbox{SINR}^{\sst {ZF}}}   
\newcommand{\rhou}{\rho_{\tiny\rm{u}}}
\newcommand{\rhod}{\rho_{\tiny\rm{d}}}

\newcommand{\bg}{{\bf{g}}}

\newcommand{\sst}{\scriptsize\mbox}   % for scriptsize text

\newcommand{\rR}{\mathbb{R}}
\newcommand{\cC}{\mathbb{C}}

\newcounter{eqncnt}

\newcounter{eqnback}

\ifCLASSINFOpdf
\else
\fi

\begin{document}
\title{Multi-Cell Massive MIMO in LoS}

\author{
\IEEEauthorblockN{Hong Yang}
\IEEEauthorblockA{\it Nokia Bell Labs, Murray Hill, NJ, USA\\
\tt{h.yang@nokia-bell-labs.com}}
\and
\IEEEauthorblockN{Hien Q. Ngo}
\IEEEauthorblockA{\it Queen's University, Belfast, UK\\
\tt{hien.ngo@qub.ac.uk}}
\and
\IEEEauthorblockN{Erik G. Larsson}
\IEEEauthorblockA{\it Link\"oping University, Sweden\\
\tt{erik.g.larsson@liu.se}}
}

\maketitle

\begin{abstract}
%\boldmath
We consider a multi-cell Massive MIMO system in a  line-of-sight (LoS) propagation environment, for    which each user is served by one base station, with no cooperation among the base stations.
Each base station knows the channel between its service antennas and its users, and uses these
channels for precoding and decoding. Under these assumptions we derive explicit downlink and uplink effective SINR formulas for maximum-ratio (MR) processing and zero-forcing (ZF) processing. We also derive formulas for power control to meet pre-determined SINR targets. A numerical example demonstrating the usage of the derived formulas is provided.
\end{abstract}

\begin{IEEEkeywords}
Massive MIMO, multi-cell, line-of-sight, power control, SINR, maximum ratio, zero-forcing.
\end{IEEEkeywords}

% \fi
\IEEEpeerreviewmaketitle

\section{Introduction}

\IEEEPARstart{T}{he} performance of Massive MIMO (multi-input/multi-output)  in rich scattering environments has been an active research area  since the seminal paper \cite{Mar:10:WCOM}. 
There is a fairly extensive collection of literature on this topic, which is expected to   grow further
with the advent of 5G deployments \cite{BSHD:15:WCOM,LLSAZ:14:JSTSP}. In comparison, the literature on Massive MIMO performance in line-of-sight (LoS) environments is rather scanty. Besides the practical consideration of deployment scenarios, this discrepancy can be attributed to the fact that the assumption of rich scattering implies that the communication channels can be reasonably modeled as independent and identically distributed Rayleigh fading, which crucially facilitates the derivation of very comprehensive and tight performance bounds for Massive MIMO \cite{MLYN:16:Book}. 

LoS channels are substantially deterministic, and therefore are in a sense the direct opposite of the rich scattering channels. In \cite{YM:17:COM}, the authors study the performance of Massive MIMO in LoS in a single-cell setting. The purpose of this paper is to extend the SINR (signal-to-interference plus noise ratio) formulas in \cite{YM:17:COM} to a multi-cell setting, and derive power control policies that meet given target SINRs. In addition to their theoretical interest, such formulas are essential for the analysis of  the physical layer throughput performance of Massive MIMO systems deployed in open spaces such as rural areas \cite{SCHPCK:12:PIRMC}, and for millimeter wave fixed wireless systems whose cell sizes are very small so that the propagation environment can be expected to be predominantly LoS \cite{BH:15:WCM}. 

We assume a general multi-cell scenario where base stations do not cooperate. Each base station serves a set of users, and each user is served by only one base station. In addition,
there is negligible mobility and we therefore  assume that each base station has perfect knowledge of the channels between each of its service antennas and each of the users that it serves.
All base stations employ either maximum-ratio (MR) or zero-forcing (ZF) precoding/decoding  for the downlink and uplink data transmissions. 

\section{Preliminaries}\label{pre}

\subsection{Notation}
Let $\rR_{+}$ denote the set of all positive real numbers and let $\rR_{0+}=\{ 0\}\cup \rR_{+}$. Let $\rR_{+}^n$, $\rR_{0+}^n$, $\rR_{+}^{m\times n}$ and $\rR_{0+}^{m\times n}$ denote the corresponding $n$-dimensional and $(m\times n)$-dimensional product spaces. Replacing $\rR$ with $\cC$ denotes the the corresponding complex spaces.

Superscripts: ${\sst T}$ denotes matrix transpose; ${\sst *}$ denotes conjugate transpose. Thus double superscripts ${\sst T*}$ and $\sst{*T}$ both denote un-transposed conjugate. 

For a vector ${\bf v}$, $[{\bf v}]_k$ denotes the $k$th element of  $\bf v$, and $\|{\bf v}\|_p$ denotes the $p$-norm.

For a matrix $A$, we define $A^{(*)} \triangleq A^*A$ to shorten the expressions. Therefore, in the case when $A$ is a column vector, we have $A^{(*)}=\|A\|_2^2$. In addition, $[A]_{n,n}$ denotes the $n$th diagonal element of $A$, and $\left| A \right|^2$ denotes element-wise magnitude square for  $A$.

Finally, $\odot$ and ${\text E} (\cdot)$ denote the Hadamard product and the  expectation operator, respectively.

\subsection{Massive MIMO System}
We consider a multi-cell Massive MIMO system that consists of $L$ cells. Each cell is served by an $M$-antenna Massive MIMO base station, which serves $K$ simultaneous users. Let
$$
G_{l'}^l=\left[ \bg_{l',1}^{l}\hspace{.1in} \cdots \hspace{.1in} \bg_{l',K}^{l} \right]\in {\mathbb C}^{M\times K}
$$
be the $M\times K$ channel matrix between an $M$-antenna array at the $l$th base station and the $K$ user terminals that are served by the $l'$th base station.

Here without loss of generality we have assumed common numbers of service antennas $M$ and simultaneous users $K$ for each cell. The scenarios of different $M$ and $K$ in different cells can be accommodated by setting some channel gain entries in the channel matrices $G_{l'}^l$ to zero. 

The downlink data channel is modeled as 
\begin{equation}
{\bf x}_l = \sqrt{\rhod} (G_l^l)^{\sst T}{\bf s}_l +  \sqrt{\rhod} \sum_{l'\ne l}(G_l^{l'})^{\sst T}{\bf s}_{l'} +{\bf w}_l,  \label{dldc}
\end{equation}
where ${\bf x}_l\in {\mathbb C}^K$ is the received signal vector at the $K$ user terminals, $\rhod$ is the normalized (with respect to noise power at the user terminal receiver) downlink signal-to-noise ratio (SNR),   ${\bf s}_l \in {\mathbb C}^M$ is the precoded input vector to the $M$-antenna ports in the $l$-th base station, and ${\bf w}_l\in {\mathbb C}^K$ is a circularly-symmetric Gaussian noise vector. Downlink power is subject to total power constraint and can be specified as 
\begin{equation}
{\text E}(\|{\bf s}_l\|_2^2)\le 1.  \label{dlpc}
\end{equation}

Similarly, denoting the corresponding uplink variables with a $\breve {\phantom{x}}$ above the variable, the uplink data channel is modeled as
\begin{equation}
{\breve{\bf x}}_l = \sqrt{\rhou} G_l^l{\breve{\bf s}}_l +\sqrt{\rhou} \sum_{l'\ne l}G_{l'}^l{\breve{\bf s}}_{l'}+ {\breve{\bf w}}_l,   \label{uldc}
\end{equation}
where ${\breve{\bf x}}_l\in {\mathbb C}^M$ is the received signal vector at the $M$-antenna ports in the $l$th base station, $\rhou$ is the normalized (with respect to noise power at the base station receiver) uplink SNR, ${\breve{\bf s}}_l, {\breve{\bf s}}_{l'}\in {\mathbb C}^K$ are the power controlled message-bearing signal from the $K$ user terminals in the $l$th cell and $l'$th cell respectively, ${\breve{\bf w}}_l\in {\mathbb C}^M$ is  a circularly-symmetric Gaussian noise vector in the $l$th cell. Uplink power is subject to individual power constraint and can be specified as
\begin{equation}
\| {\text E} ({\breve{\bf s}}^{*\sst T}_l\odot {\breve{\bf s}}_l) \|_\infty\le 1,  \label{ulpc}
\end{equation}
where $\odot$ denotes the element-wise multiplication.

\section{Effective SINR}\label{s:sinr}
In LoS environment, channels are substantially constant. In this section, we derive both downlink and uplink SINR expressions in terms of the channel matrix $G_{l'}^l$ for a multi cell Massive MIMO system with maximum-ratio processing and zero-forcing processing.

In the following derivations of effective SINR, the coding is performed over many random realizations of the user message-bearing symbols ${\bf q}_l$ and the noise ${\bf w}_l$, ${\breve{\bf w}}_l$,  but the channel $G_{l'}^l$ is assumed to be constant, i.e., the expectation ${\text E}$ is taken only with respect to  ${\bf q}_l$, ${\breve{\bf q}}_l$ and ${\bf w}_l$, ${\breve{\bf w}}_l$.
We shall assume that the user data symbols ${\bf q}_l$ and ${\breve{\bf q}}_l$ have zero mean and unit variance, and uncorrelated, and user data from different cells are uncorrelated, that is
\begin{equation}
{\text E}({\bf q}_l{\bf q}_{l'}^*) =\delta_{l, l'} I_K, \quad  {\text E}({\breve{\bf q}}_l{\breve{\bf q}}^*_{l'}) =\delta_{l, l'} I_K\label{eqq},
\end{equation}
here $\delta_{l, l'}$ is the Kronecker delta, and Gaussian noise with
\begin{equation}
{\text E}({\bf w}_l{\bf w}_l^*) = I_K  \quad  {\text E}({\breve{\bf w}}_l{\breve{\bf w}}^*_l) = I_M \label{eww},
\end{equation}
where $I_K$ and $I_M$ are $K$ dimensional and $M$ dimensional identity matrices. 

\subsection{Maximum-Ratio Downlink SINR}
We use the precoding matrix 
$$
(G_l^l)^{{\sst T}*}D_{(G_l^l)^{(*)}}^{-1/2}D_{\eta_l}^{1/2},
$$
where the diagonal matrices
\begin{eqnarray}
D_{(G_l^l)^{(*)}}^{-1/2} &&\hspace{-.25in}\triangleq \mbox{diag}\left( [(G_l^l)^{(*)}]_{1,1}^{-1/2}, \cdots, [(G_l^l)^{(*)}]_{K,K}^{-1/2}\right) \nonumber \\
&&\hspace{-.25in}= \mbox{diag}\left( \|\bg_{l,1}^l\|_2^{-1}, \cdots, \|\bg_{l,K}^l\|_2^{-1}\right), \label{eq:dzz}
\end{eqnarray}
and
\begin{equation}
D_{\eta_l}^{1/2} \triangleq \mbox{diag}\left(\sqrt{\eta_{l,1}}, \cdots, \sqrt{\eta_{l,K}}\right). \label{eq:deta}
\end{equation}
Here $\boldsymbol\eta_l =(\eta_{l,1} \hspace{.1in} \cdots \hspace{.1in}\eta_{l,K})^{\sst T}$ is the downlink power control. $\boldsymbol\eta_l$ must satisfy the total power constraint
\begin{equation}\label{dlpc_c}
\boldsymbol\eta_l \in {\cal A}_1\triangleq \{{\bf u}\in {\mathbb R}_{0+}^{K}: \|{\bf u}\|_1 \le 1\}.
\end{equation}

We have
\begin{equation}
{\bf s}_l = (G_l^l)^{{\sst T}*}D_{(G_l^l)^{(*)}}^{-1/2}D_{\eta_l}^{1/2}{\bf q}_l,  \label{mr_d}
\end{equation}
where ${\bf q}_l\in {\mathbb C}^K$ is the message-bearing vector.

Using (\ref{eqq}) we see that  (\ref{dlpc}) is satisfied:
\begin{eqnarray*}
\lefteqn{{\text E}({\bf s}_l^{(*)}) } \\
&& =\mbox{tr}\left[((G_l^l)^{(*)})^{\sst T}\mbox{diag}\left( \|\bg_{l,1}^l\|_2^{-2}\eta_{l,1}, \cdots, \|\bg_{l,K}^l\|_2^{-2}\eta_{l,K}\right)\right] \\
&& =\|\boldsymbol\eta_l\|_1\le 1.
\end{eqnarray*}

We have
\begin{eqnarray}
\lefteqn{{\bf x}_l = \sqrt{\rhod} (G_l^l)^{\sst T} (G_l^l)^{{\sst T}*}D_{(G_l^l)^{(*)}}^{-1/2}D_{\eta_l}^{1/2}{\bf q}_l} \nonumber \\
&& \hspace{1.0in}  + \sqrt{\rhod} \sum_{l'\ne l}(G_l^{l'})^{\sst T}{\bf s}_{l'}+ {\bf w}_l.
\end{eqnarray}

The $k$th user terminal in the $l$th cell receives 
\begin{eqnarray}
x_{l,k} &&\hspace{-.25in}= \sqrt{\rhod} (\bg_{l,k}^l)^{\sst T}{(\bg_{l,k}^l)^{{\sst T}*}\over \|\bg_{l,k}^l\|_2}\eta_{l,k}^{1/2}q_{l,k} + \left(w_{l,k} + \phantom{\sum_{l'\ne l}(\bg_{l,k}^{l'})^{\sst T}{\bf s}_{l'}} \right.  \nonumber \\
&& \hspace{.5in} \sqrt{\rhod} (\bg_{l,k}^l)^{\sst T}\sum_{k'\ne k}{(\bg_{l,k'}^l)^{{\sst T}*}\over \|\bg_{l,k'}^l\|_2}\eta_{l,k'}^{1/2}q_{l,k'} +\nonumber \\
&& \hspace{1in}\left.   \sqrt{\rhod} \sum_{l'\ne l}(\bg_{l,k}^{l'})^{\sst T}{\bf s}_{l'}\right) \nonumber\\
&&\hspace{-.25in}= \sqrt{\rhod\eta_{l,k}} \|\bg_{l,k}^l\|_2q_{l,k} + \left(w_{l,k} + \phantom{\sum_{l'\ne l}(\bg_{l,k}^{l'})^{\sst T}{\bf s}_{l'}} \right.  \nonumber \\
&& \hspace{.5in}\sqrt{\rhod} (\bg_{l,k}^l)^{\sst T}\sum_{k'\ne k}{(\bg_{l,k'}^l)^{{\sst T}*}\over \|\bg_{l,k'}^l\|_2}\eta_{l,k'}^{1/2}q_{l,k'}+ \nonumber \\
&& \hspace{1in} \left. \sqrt{\rhod} \sum_{l'\ne l}(\bg_{l,k}^{l'})^{\sst T}{\bf s}_{l'}\right).
 \label{mrdxk}
\end{eqnarray}

The right-hand-side of (\ref{mrdxk}) has four terms. We compute the power of each term in the following.
\begin{itemize}
\item Signal power (SP):
$$
 \rhod\eta_{l,k}\|\bg_{l,k}^l\|_2^2{\text E}(q_{l,k}^*q_{l,k}) =  \rhod\eta_{l,k}\|\bg_{l,k}^l\|_2^2. \\
$$

\item Noise power (NP): 
\begin{equation}
{\text E}(w_{l,k}^*w_{l,k}) = 1.  \label{dnp}
\end{equation}

\item Interference power from MR processing (IP):
\begin{eqnarray*}
\lefteqn{\rhod{\text E}\left[\left((\bg_{l,k}^l)^{\sst T}\sum_{k'\ne k}{(\bg_{l,k'}^l)^{{\sst T}*}\over \|\bg_{l,k'}^l\|_2}\eta_{l,k'}^{1/2}q_{l,k'}\right)^{(*)}\right]} \hspace{.2in} \\
&& = \rhod \sum_{k'\ne k} \eta_{l,k'}{(\bg_{l,k'}^l)^{\sst T}(\bg_{l,k}^l)^{{\sst T}*}(\bg_{l,k}^l)^{\sst T}(\bg_{l,k'}^l)^{{\sst T}*}\over \|\bg_{l,k'}^l\|_2^2} \\
&&= \rhod\sum_{k'\ne k} \eta_{l,k'}{|(\bg_{l,k}^l)^*\bg_{l,k'}^l|^2\over \|\bg_{l,k'}^l\|_2^2}.
\end{eqnarray*}

\item Interference power from base stations in other cells (OP):

Using (\ref{eqq}) and (\ref{mr_d}), we have 
\begin{eqnarray*}
\lefteqn{\rhod{\text E}\left[ \left(\sum_{l'\ne l}(\bg_{l,k}^{l'})^{\sst T}{\bf s}_{l'}\right)^{(*)} \right]= } \\
&&\hspace{1in}\rhod\sum_{l'\ne l}\sum_{k'=1}^K \eta_{l',k'}{|(\bg_{l,k}^{l'})^*\bg_{l',k'}^{l'}|^2\over \|\bg_{l',k'}^{l'}\|_2^2}.
\end{eqnarray*}
\end{itemize}

From the above calculations, we obtain the SINR given by \eqref{sinr_mr_dl}, shown at the top of the next page.

%%%%%%%%%%%%%%%%%%%%%%%%%%%% Equation (15) %%%%%%%%%%%%%%%%%%%%%%%%%%%%%%%%%%%%%%
\setcounter{eqnback}{\value{equation}} \setcounter{equation}{13}
\begin{figure*}[!t]

\begin{eqnarray}
 \SMR_{l,k}={\mbox{SP}\over \mbox{NP} + \mbox{IP}+\mbox{OP}}= 
{\rhod\eta_{l,k}\|\bg_{l,k}^l\|_2^2 \over 1+\displaystyle\rhod\sum_{k'\ne k} \eta_{l,k'}{|(\bg_{l,k}^l)^*\bg_{l,k'}^l|^2\over \|\bg_{l,k'}^l\|_2^2}+\rhod\sum_{l'\ne l}\sum_{k'=1}^K \eta_{l',k'}{|(\bg_{l,k}^{l'})^*\bg_{l',k'}^{l'}|^2\over \|\bg_{l',k'}^{l'}\|_2^2}}, \label{sinr_mr_dl}
\end{eqnarray}

\hrulefill
\end{figure*}
\setcounter{eqncnt}{\value{equation}}
\setcounter{equation}{\value{eqnback}}
%%%%%%%%%%%%%%%%%%%%%%%%%%%%%%%%%%%%%%%%%%%%%%%%%%%%%%%%%%%%%%%%%%%%%%%%%%%%%%%%%%%%

To achieve this SINR, the base station needs to know the denominator of the SINR. Though this quantity depends on the channels of other users in other cells, the base station does not require knowledge of all channels. In practice, this quantity can be achieved by estimating the power of the aggregate interference.

\subsection{Maximum-Ratio Uplink SINR}
For MR decoding, the decoding matrix is $(G_l^l)^*$. From (\ref{uldc}) we have
$$
(G_l^l)^*{\breve{\bf x}}_l = \sqrt{\rhou}(G_l^l)^{(*)}{\breve{\bf s}}_l +\sqrt{\rhou}\sum_{l'\ne l}(G_l^l)^* G_{l'}^l{\breve{\bf s}}_{l'} +(G_l^l)^*{\breve{\bf w}}_l,   
$$
where ${\breve{\bf s}}_l=D_{\eta_l}^{1/2}{\breve{\bf q}}_l$ is the power controlled message-bearing signal vector from the $K$ user terminals in the $l$th cell. The uplink power control $\breve{\boldsymbol\eta}_l =(\breve{\eta}_{l,1} \hspace{.1in} \cdots \hspace{.1in}\breve{\eta}_{l,K})^{\sst T}$ must satisfy the individual power constraint
\setcounter{equation}{14}\begin{equation}\label{ulpc_c}
\breve{\boldsymbol\eta}_l \in {\cal A}_\infty \triangleq \{{\bf u}\in {\mathbb R}_{0+}^{K}: \|{\bf u}\|_\infty \le 1\}.
\end{equation}

For the $k$th user terminal in the $l$th cell
\begin{eqnarray}
[(G_l^l)^*{\breve{\bf x}}_l]_k && \hspace{-.25in}= \sqrt{\rhou}(\bg_{l,k}^l)^* G_l^l{\breve{\bf s}}_l + \nonumber\\
&&\hspace{-.5in} \sqrt{\rhou}\sum_{l'\ne l}(\bg_{l,k}^l)^* G_{l'}^l{\breve{\bf s}}_{l'}+(\bg_{l,k}^l)^*{\breve{\bf w}}_l  \nonumber \\
&& \hspace{-.25in}=\sqrt{\rhou\breve{\eta}_{l,k}}(\bg_{l,k}^l)^*\bg_{l,k}^lq_{l,k}' +(\bg_{l,k}^l)^*\left[{\breve{\bf w}}_l + \phantom{\sum_{l'\ne l}} \right. \nonumber\\
&& \hspace{-.5in}\left. \sqrt{\rhou}\sum_{k'\ne k}\sqrt{\breve{\eta}_{l,k'}}\bg_{l,k'}^lq_{l,k'}'+\sqrt{\rhou}\sum_{l'\ne l}G_{l'}^l{\breve{\bf s}}_{l'}\right]. \label{mruxk}
\end{eqnarray}

Similar to the downlink case, there are four terms in (\ref{mruxk}), 
\begin{itemize}
\item Signal power ($\breve{\mbox{SP}}$):
$$
\rhou\breve{\eta}_{l,k}\|\bg_{l,k}^l\|_2^4.
$$

\item Noise power ($\breve{\mbox{NP}}$): 
$$
\|\bg_{l,k}^l\|_2^2.
$$

\item Interference power from MR processing ($\breve{\mbox{IP}}$):
\begin{eqnarray*}
\lefteqn{\rhou {\text E}\left[\left((\bg_{l,k}^l)^*\sum_{k'\ne k}\sqrt{\breve{\eta}_{l,k'}}\bg_{l,k'}^lq'_{l,k'}\right)^{(*)}\right]} \\
&&= {\rhou} \sum_{k'\ne k} \breve{\eta}_{l,k'}(\bg_{l,k'}^l)^*\bg_{l,k}^l(\bg_{l,k}^l)^*\bg_{l,k'}^l \\
&&= {\rhou} \sum_{k'\ne k} \breve{\eta}_{l,k'}|(\bg_{l,k}^l)^*\bg_{l,k'}^l |^2.
\end{eqnarray*} 

\item Interference power from users in other cells ($\breve{\mbox{OP}}$):
\begin{eqnarray*}
\lefteqn{\rhou {\text E}\left[\left((\bg_{l,k}^l)^*\sum_{l'\ne l}G_{l'}^l{\breve{\bf s}}_{l'}\right)^{(*)} \right]} \\
&&= {\rhou} \sum_{l'\ne l}\sum_{k'=1}^K\breve{\eta}_{l',k'}(\bg_{l',k'}^l)^*\bg_{l,k}^l(\bg_{l,k}^l)^*\bg_{l',k'}^l\\
&&={\rhou} \sum_{l'\ne l}\sum_{k'=1}^K\breve{\eta}_{l',k'}|(\bg_{l,k}^l)^*\bg_{l',k'}^l|^2.
\end{eqnarray*} 
\end{itemize}
Thus, we can obtain the corresponding SINR given by \eqref{sinr_mr_ul}, shown at the top of the next page.

%%%%%%%%%%%%%%%%%%%%%%%%%%%% Equation (15) %%%%%%%%%%%%%%%%%%%%%%%%%%%%%%%%%%%%%%
\setcounter{eqnback}{\value{equation}} \setcounter{equation}{16}
\begin{figure*}[!t]
\begin{eqnarray}
\SMR_{l,k}={\breve{\mbox{SP}}\over \breve{\mbox{NP}} + \breve{\mbox{IP}}+ \breve{\mbox{OP}}}=  
{\|\bg_{l,k}^l\|_2^2\rhou\breve{\eta}_{l,k}\over 1+\displaystyle{\rhou\over \|\bg_{l,k}^l\|_2^2}\left(\sum_{k'\ne k} \breve{\eta}_{l,k'}|(\bg_{l,k}^l)^*\bg_{l,k'}^l |^2 +\sum_{l'\ne l}\sum_{k'=1}^K\breve{\eta}_{l',k'}|(\bg_{l,k}^l)^*\bg_{l',k'}^l|^2 \right)}, \label{sinr_mr_ul}
\end{eqnarray}
\hrulefill
\end{figure*}
\setcounter{eqncnt}{\value{equation}}
\setcounter{equation}{\value{eqnback}}
%%%%%%%%%%%%%%%%%%%%%%%%%%%%%%%%%%%%%%%%%%%%%%%%%%%%%%%%%%%%%%%%%%%%%%%%%%%%%%%%%%%%

\subsection{Zero-Forcing Downlink SINR}
For zero-forcing precoding we use the precoding matrix
$$
 (G_l^l)^{\sst T*}((G_l^l)^{\sst T}(G_l^l)^{\sst T*})^{-1}D_{((G_l^l)^{(*)})^{-1}}^{-1/2}D_{\eta_l}^{1/2},
$$
where
\begin{eqnarray*}
\lefteqn{D_{((G_l^l)^{(*)})^{-1}}^{-1/2}\triangleq }  \\
&&\mbox{diag}\left( [((G_l^l)^{(*)})^{-1}]_{1,1}^{-1/2}, \cdots, [((G_l^l)^{(*)})^{-1}]_{K,K}^{-1/2}\right),
\end{eqnarray*}
and $D_{\eta_l}^{1/2}$ is defined by (\ref{eq:deta}).

We have
\setcounter{equation}{17}\begin{equation}
{\bf s}_l = (G_l^l)^{\sst T*}((G_l^l)^{\sst T}(G_l^l)^{\sst T*})^{-1}D_{((G_l^l)^{(*)})^{-1}}^{-1/2}D_{\eta_l}^{1/2}{\bf q}_l.  \label{zf_d}
\end{equation}

With the assumption (\ref{eqq}), we calculate that 
\begin{eqnarray*}
\lefteqn{{\text E}({\bf s}_l^{(*)}) } \\
&& =\mbox{tr}\left[D_{((G_l^l)^{(*)})^{-1}}^{-1/2}(((G_l^l)^{(*)})^{-1})^{\sst T}D_{((G_l^l)^{(*)})^{-1}}^{-1/2}D_{\eta_l}\right] \\
&& =\|\boldsymbol\eta_l\|_1\le 1,
\end{eqnarray*}
which satisfies the downlink power constraint (\ref{dlpc}).

Thus,
\begin{eqnarray}
{\bf x}_l && \hspace{-.25in} = \sqrt{\rhod} (G_l^l)^{\sst T}(G_l^l)^{\sst T*}((G_l^l)^{\sst T}(G_l^l)^{\sst T*})^{-1}D_{((G_l^l)^{(*)})^{-1}}^{-1/2}D_{\eta_l}^{1/2}{\bf q}_l \nonumber\\ 
&& + \sqrt{\rhod}\sum_{l'\ne l}(G_l^{l'})^{\sst T}{\bf s}_{l'}+ {\bf w}_l.
\end{eqnarray}
The $k$th user terminal  in the $l$th cell receives
\begin{equation}
x_{l,k} = \sqrt{\rhod\eta_{l,k}\over [((G_l^l)^{(*)})^{-1}]_{k,k}}q_{l,k} + \sqrt{\rhod}\sum_{l'\ne l}(\bg_{l,k}^{l'})^{\sst T}{\bf s}_{l'} +w_{l,k}.  \label{zfdxk}
\end{equation}

The signal power is calculated as
$$
{\rhod\eta_{l,k}\over [((G_l^l)^{(*)})^{-1}]_{k,k}}{\text E}(q_{l,k}^{(*)}) = {\rhod\eta_{l,k}\over [((G_l^l)^{(*)})^{-1}]_{k,k}}.
$$

The noise power is 1 as in (\ref{dnp}). 

From (\ref{zf_d}) and (\ref{eqq}), and letting
\begin{eqnarray*}
\lefteqn{{\bf a}_{l,k}^{l'} }\\
&& \hspace{-.25in} =\left[ (\bg_{l,k}^{l'})^{\sst T}(G_{l'}^{l'})^{\sst T*}((G_{l'}^{l'})^{\sst T}(G_{l'}^{l'})^{\sst T*})^{-1}D_{((G_{l'}^{l'})^{(*)})^{-1}}^{-1/2}D_{\eta_{l'}}^{1/2}\right]^{\sst T} \\
&& \hspace{-.25in} =D_{\eta_{l'}}^{1/2}D_{((G_{l'}^{l'})^{(*)})^{-1}}^{-1/2}((G_{l'}^{l'})^{(*)})^{-1}(G_{l'}^{l'})^{*}\bg_{l,k}^{l'},
\end{eqnarray*}•
the interference power from base stations in other cells is
\begin{eqnarray*}
&&\hspace{-.25in}\mbox{OP}_{l,k}^{\sst {ZF,d}}=\rhod{\text E}\left[\left(\sum_{l'\ne l}(\bg_{l,k}^{l'})^{\sst T}{\bf s}_{l'}\right)^{(*)}\right] =\rhod \sum_{l'\ne l}({\bf a}_{l,k}^{l'})^{(*)}\\
%=\rhod \sum_{l'\ne l}(\bg_{l,k}^{l'})^*G_{l'}^{l'}((G_{l'}^{l'})^{(*)})^{-1}D_{((G_{l'}^{l'})^{(*)})^{-1}}^{-1}D_{\eta_{l'}}((G_{l'}^{l'})^{(*)})^{-1}(G_{l'}^{l'})^{*}\bg_{l,k}^{l'}\\
&&\hspace{-.25in}=\rhod \sum_{l'\ne l}\mbox{tr}\left[ \left( (\bg_{l,k}^{l'})^*G_{l'}^{l'}((G_{l'}^{l'})^{(*)})^{-1}\right)^{(*)}D_{((G_{l'}^{l'})^{(*)})^{-1}}^{-1}D_{\eta_{l'}}\right] \\
&&\hspace{-.25in}=\rhod \sum_{l'\ne l}\sum_{k'=1}^K {\left[ \left((\bg_{l,k}^{l'})^*G_{l'}^{l'}((G_{l'}^{l'})^{(*)})^{-1}\right)^{(*)}\right]_{k',k'}\over \left[ ((G_{l'}^{l'})^{(*)})^{-1}\right]_{k',k'}} \eta_{l',k'}. 
\end{eqnarray*}

Thus, 
$$
\SZF_{l,k}={\rhod\eta_{l,k}\over (1+ \mbox{OP}_{l,k}^{\sst {ZF,d}})[((G_l^l)^{(*)})^{-1}]_{k,k}}.
$$

\subsection{Zero-Forcing Uplink SINR}
For ZF decoding, the decoding matrix for the base station in the $l$th cell is $((G_l^l)^{(*)})^{-1}(G_l^l)^*$. From (\ref{uldc}) we have
\begin{eqnarray*}
\lefteqn{((G_l^l)^{(*)})^{-1}(G_l^l)^*{\breve{\bf x}}_l = \sqrt{\rhou}((G_l^l)^{(*)})^{-1}(G_l^l)^{(*)}{\breve{\bf s}}_l +}\\
&& \sqrt{\rhou}((G_l^l)^{(*)})^{-1}(G_l^l)^* \sum_{l'\ne l}G_{l'}^l{\breve{\bf s}}_{l'}  + ((G_l^l)^{(*)})^{-1}(G_l^l)^*{\breve{\bf w}}_l,   
\end{eqnarray*}
where ${\breve{\bf s}}_l=D_{\breve{\eta}_l}^{1/2}{\breve{\bf q}}_l$ is the power controlled message-bearing signal vector from the $K$ user terminals in the $l$th cell.  

For the $k$th user terminal in the $l$th cell,
\begin{eqnarray}
\lefteqn{ \left[((G_l^l)^{(*)})^{-1}(G_l^l)^*{\breve{\bf x}}_l\right]_k = \sqrt{\rhou\breve{\eta}_{l,k}}q'_{l,k} + }\nonumber\\
&& \hspace{.5in}\sqrt{\rhou} \sum_{l'\ne l}\left[((G_l^l)^{(*)})^{-1}(G_l^l)^*G_{l'}^l{\breve{\bf s}}_{l'}\right]_k  +  \nonumber\\
 && \hspace{1in} \left[((G_l^l)^{(*)})^{-1}(G_l^l)^*{\breve{\bf w}}_l\right]_k.    \label{zfuxk}
\end{eqnarray}

The signal power is calculated as $\rhou\breve{\eta}_{l,k}$. 
The noise power is
$$
{\text E}\left[((G_l^l)^{(*)})^{-1}(G_l^l)^*{\breve{\bf w}}_l\right]^{(*)}_k 
$$
which is the $k$th diagonal element of the covariance matrix
\begin{eqnarray*}
\lefteqn{{\mbox{Cov}}[((G_l^l)^{(*)})^{-1}(G_l^l)^*{\breve{\bf w}}_l]}\\
&&={\text E}\left[((G_l^l)^{(*)})^{-1}(G_l^l)^*{\breve{\bf w}}_l(((G_l^l)^{(*)})^{-1}(G_l^l)^*{\breve{\bf w}}_l)^* \right]\\
&&={\text E}\left[((G_l^l)^{(*)})^{-1}(G_l^l)^*{\breve{\bf w}}_l{\breve{\bf w}}^*_lG_l^l((G_l^l)^{(*)})^{-1}\right]\\
&&=((G_l^l)^{(*)})^{-1}(G_l^l)^*{\text E}\left[{\breve{\bf w}}_l{\breve{\bf w}}^*_l\right]G_l^l((G_l^l)^{(*)})^{-1}\\
&&=((G_l^l)^{(*)})^{-1}(G_l^l)^*I_M G_l^l((G_l^l)^{(*)})^{-1}\\
&&=((G_l^l)^{(*)})^{-1}.
\end{eqnarray*}
Thus, the noise power 
$$
{\text E}\left[((G_l^l)^{(*)})^{-1}(G_l^l)^*{\breve{\bf w}}_l\right]^{(*)}_k = [((G_l^l)^{(*)})^{-1}]_{k,k}.
$$

The interference power from users in other cells is
$$
\mbox{OP}_{l,k}^{\sst {ZF,u}}=\rhou {\text E}\left[\sum_{l'\ne l}B_{l'}^l{\breve{\bf s}}_{l'}\right]_k^{(*)},
$$
where
$$
B_{l'}^l=\left((G_l^l)^{(*)}\right)^{-1}(G_l^l)^*G_{l'}^l.
$$

Then using (\ref{eqq}) we have
\begin{eqnarray*}
\mbox{OP}_{l,k}^{\sst {ZF,u}}&&\hspace{-.25in}=\rhou \sum_{l'\ne l} \left[(B_{l'}^l)^* D_{\breve{\eta}_{l'}}B_{l'}^l\right]_{k,k} \\
&&\hspace{-.25in}=\rhou \sum_{l'\ne l} \sum_{k'=1}^K \left[B_{l'}^l\right]_{k,k'}^{(*)}\breve{\eta}_{l',k'}.
\end{eqnarray*}

Therefore, we obtain
$$
\SZF_{l,k}={\rhou\breve{\eta}_{l,k}\over [((G_l^l)^{(*)})^{-1}]_{k,k}+\rhou \displaystyle\sum_{l'\ne l} \sum_{k'=1}^K [B_{l'}^l]_{k,k'}^{(*)}\breve{\eta}_{l',k'} }.
$$

\section{Power Control}
Based on the SINR expressions obtained in Section \ref{s:sinr}, power controls to achieve given SINR targets can be readily obtained. 

For a set of $KL$ given SINR target 
$$
\left\{ \zeta_{l,k}: l=1, \cdots, L, k=1, \cdots, K\right\},
$$
write 
\begin{equation}\label{eq:tsinr}
{\boldsymbol\zeta}=\left [\zeta_{1,1},\cdots \zeta_{1,K},\cdots, \zeta_{L,1},\cdots, \zeta_{L,K}\right]^{\sst T}\in {\mathbb R}_{0+}^{KL} .
\end{equation}•
and 
$$
D_{\boldsymbol\zeta}\triangleq  \mbox{diag}(\boldsymbol\zeta)\in {\mathbb R}_{0+}^{KL\times KL}.
$$
Then by the SINR expressions derived in Section \ref{s:sinr}, target SINR given by (\ref{eq:tsinr}) can be met if and only if the system of linear equations 
$$
(D-D_{\boldsymbol\zeta}C)\boldsymbol\eta =\boldsymbol\zeta
$$
has a solution $\boldsymbol\eta \in {\mathbb R}_{0+}^{KL}$ and $\|\boldsymbol\eta_l\|_e \le 1, l=1, \cdots \, L$, where $e=1$ for downlink and $e=\infty$ for uplink. 

We list the corresponding diagonal matrix $D$ and matrix $C$ for MR downlink, MR uplink, ZF downlink, and ZF uplink in the following subsections.

\subsection{Maximum-Ratio Downlink Power Control}\label{s:pcmrd}
Let 
\begin{eqnarray}
{\bf v}_l^{\sst{MR}}&&\hspace{-.25in}\triangleq \left( \|\bg_{l,1}^l\|_2^2, \cdots, \|\bg_{l,K}^l\|_2^2\right),\quad  l=1, \cdots, L, \nonumber \\
{\bf v}^{\sst{MR}}&&\hspace{-.25in}\triangleq\left( {\bf v}_1^{\sst{MR}}, \cdots, {\bf v}_L^{\sst{MR}}\right), \hspace{.3in} \label{eq:vmr}
\end{eqnarray}
and
$$
C_{\sst d}^{\sst{MR}}\triangleq \left[
\begin{array}{cccc}
(G_1^1)^{(*)} & (G_1^2)^*G_2^2 & \cdots & (G_1^L)^*G_L^L \\
\vdots & \vdots & \vdots & \vdots \\
(G_L^1)^*G_1^1 & (G_L^2)^*G_2^2 & \cdots & (G_L^L)^{(*)}
\end{array}•
\right].
$$
Then
\begin{eqnarray*}
D&&\hspace{-.25in}=\rhod\mbox{diag}({\bf v}^{\sst{MR}})\in {\mathbb C}^{KL\times KL}, \\
C&&\hspace{-.25in}=\rhod \left|C_{\sst d}^{\sst{MR}}-\mbox{diag}({\bf v}^{\sst{MR}})\right|^2\left[\mbox{diag}({\bf v}^{\sst{MR}})\right]^{-1}\in {\mathbb C}^{KL\times KL}.
\end{eqnarray*}•

\subsection{Maximum-Ratio Uplink Power Control}\label{s:pcmru}
With the same ${\bf v}^{\sst{MR}}$ as defined in (\ref{eq:vmr}), let 
$$
C_{\sst u}^{\sst{MR}}\triangleq \left[
\begin{array}{cccc}
(G_1^1)^{(*)} & (G_1^1)^*G_2^1 & \cdots & (G_1^1)^*G_L^1 \\
\vdots & \vdots & \vdots & \vdots \\
(G_L^L)^*G_1^L & (G_L^L)^*G_2^L & \cdots & (G_L^L)^{(*)}
\end{array}•
\right].
$$
Then
\begin{eqnarray*}
D&&\hspace{-.25in}=\rhou\mbox{diag}({\bf v}^{\sst{MR}})\in {\mathbb C}^{KL\times KL}, \\
C&&\hspace{-.25in}=\rhou \left[\mbox{diag}({\bf v}^{\sst{MR}})\right]^{-1}\left|C_{\sst u}^{\sst{MR}}-\mbox{diag}({\bf v}^{\sst{MR}})\right|^2\in {\mathbb C}^{KL\times KL}.
\end{eqnarray*}•

\subsection{Zero-Forcing Downlink Power Control}\label{s:pczfd}
Let 
\begin{eqnarray}
{\bf v}_l^{\sst{ZF}}&&\hspace{-.25in}\triangleq \left( {1\over [((G_l^l)^{(*)})^{-1}]_{1,1}}, \cdots, {1\over [((G_l^l)^{(*)})^{-1}]_{K,K}}\right), \nonumber\\
&& \hspace{2in}  l=1, \cdots, L, \nonumber \\
{\bf v}^{\sst{ZF}}&&\hspace{-.25in}\triangleq \left( {\bf v}_1^{\sst{ZF}}, \cdots, {\bf v}_L^{\sst{ZF}}\right). \label{eq:vzf}
\end{eqnarray}
$$
C_{\sst d}^{\sst{ZF}}=\left[
\begin{array}{cccc}
{\bf 0}_{K\times K} & \left| \left( B_1^2\right)^* \right|^2 & \cdots & \left| \left( B_1^L\right)^* \right|^2\\
\vdots & \vdots & \vdots & \vdots \\
\left| \left( B_L^1\right)^* \right|^2 & \left| \left( B_L^2\right)^* \right|^2 & \cdots & {\bf 0}_{K\times K}
\end{array}•
\right].
$$
Then
\begin{eqnarray*}
D&&\hspace{-.25in}=\rhod\mbox{diag}({\bf v}^{\sst{ZF}})\in {\mathbb C}^{KL\times KL}, \\
C&&\hspace{-.25in}=\rhod C_{\sst d}^{\sst{ZF}}\mbox{diag}({\bf v}^{\sst{ZF}})\in {\mathbb C}^{KL\times KL}.
\end{eqnarray*}•

\subsection{Zero-Forcing Uplink Power Control}\label{s:pczfu}
With the same ${\bf v}^{\sst{ZF}}$ as defined in (\ref{eq:vzf}), let 
$$
C_{\sst u}^{\sst{ZF}}=\left[
\begin{array}{cccc}
{\bf 0}_{K\times K}  & \left| B_2^1\right|^2 & \cdots &  \left| B_L^1\right|^2 \\
\vdots & \vdots & \vdots & \vdots \\
 \left| B_1^L\right|^2 &  \left| B_2^L\right|^2 & \cdots & {\bf 0}_{K\times K} 
\end{array}•
\right].
$$
Then
\begin{eqnarray*}
D&&\hspace{-.25in}=\rhou\mbox{diag}({\bf v}^{\sst{ZF}})\in {\mathbb C}^{KL\times KL}, \\
C&&\hspace{-.25in}=\rhou \mbox{diag}({\bf v}^{\sst{ZF}})C_{\sst u}^{\sst{ZF}}\in {\mathbb C}^{KL\times KL}.
\end{eqnarray*}•

\subsection{Single Cell Power Control}\label{s:sczf}
The max-min power control for ZF is particularly simple in single cell case. The downlink power control is given by \cite{YM:17:COM}
$$
\eta_{l,k}={\left[\left((G_l^l)^{(*)}\right)^{-1}\right]_{k,k}\over \displaystyle\sum_{k'=1}^K\left[\left((G_l^l)^{(*)}\right)^{-1}\right]_{k',k'}}.
$$
The uplink power control is given by
$$
\breve{\eta}_{l,k}={\left[ \left((G_l^l)^{(*)}\right)^{-1}\right]_{k,k}\over \displaystyle\max_{k'}\left\{ \left[ \left((G_l^l)^{(*)}\right)^{-1}\right]_{k',k'}\right\} }.
$$

The max-min power control for MR in single cell case can be obtained by testing the solvability of a linear system in a bisection search   \cite{YM:17:COM}.  

\section{Numerical Example}
In this section, we present a numerical example to demonstrate the usage of the formulas derived in the previous sections, and to show the SINR performance of a multi-cell cluster in 60 GHz band. 

\subsection{Propagation Model}
With LoS propagation, according to the spherical wave model, the channel vector for the $k$th user in the $l$th cell and the $M$-antenna array at the $l'$th cell is given by
$$
\bg_{l,k}^{l'}=\left( {e^{i{2\pi\over \lambda} r_{l,k}^{l',1}}\over r_{l,k}^{l',1}}, \cdots, {e^{i{2\pi\over \lambda} r_{l,k}^{l',M}}\over r_{l,k}^{l',M}}\right)^{\sst T}
$$
where $r_{l,k}^{l',m}$ is the distance between the $k$th user in the $l$th cell and the $m$-antenna at the $l'$th cell, $\lambda$ is the wavelength of the carrier. 

Free space path loss is given by 
$$
\mbox{PL}_{\sst{free space}}= \left( {4\pi d f\over c}\right)^2
$$
where $f$ is the carrier frequency in GHz, $d$ is the distance from the transmitter in meters, and $c=299792458$ m/s is the speed of light. In dB, we have 
$$
\mbox{PL}_{\sst{free space, dB}}=32.45 + 20\log_{10}(f)+20\log_{10}(d).
$$

\begin{figure}
\centering
\includegraphics[width=3.5in]{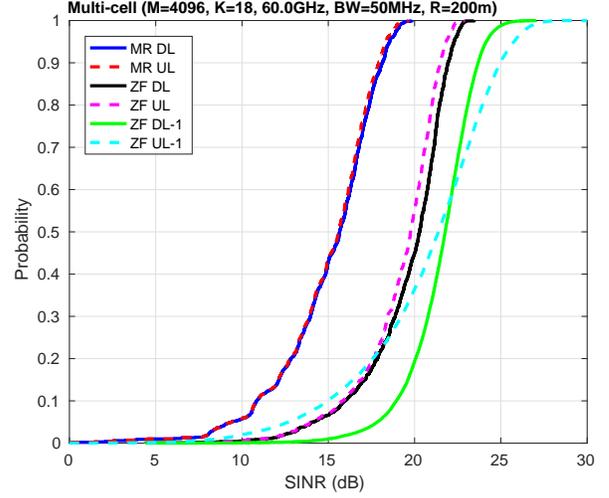}
\caption{Cumulative distribution function (CDF) of user SINR for a multi-cell Massive MIMO in LoS: ``MR DL'', ``MR UL'',  ``ZF DL'' and ``ZF UL''are the downlink and uplink system-wide max-min user SINR obtained by power controls given in Sections \ref{s:pcmrd}, \ref{s:pcmru}, \ref{s:pczfd} and \ref{s:pczfu}. ``ZF DL-1'' and ``ZF UL-1'' are the downlink and uplink user SINR in the center cell using single cell max-min power control given in Section \ref{s:sczf}.}
\label{f:sinr}
\end{figure}

In our multi-cell example, we assume a standard 7-cell cluster: a center hexagonal cell is surrounded by 6 other equal-sized hexagonal cells. Within each cell $K$ users are randomly distributed. Simulation parameters are summarized in Table \ref{t:para}. Simulation results are shown in Figure \ref{f:sinr}.
%The performance statistics is collected from the center cell.

\begin{table}
\centering
\caption{Simulation Parameters}\label{t:para}
\begin{tabular}{|l|} \hline
Number of service antennas per cell $M$ = 4096\\ \hline
Array configuration: circular, ${\lambda\over 2}$ arc separation\\ \hline
Number of simultaneous user per cell $K$ = 18  \\ \hline
Massive MIMO service antenna gain =  0 dBi \\ \hline
Mobile antenna gain = 0 dBi \\ \hline
Base station receiver noise figure = 9 dB \\ \hline
Mobile receiver noise figure = 9 dB \\ \hline
Base station radiated power = 2 W per base station\\ \hline
Mobile radiated power = 200 mW per mobile\\ \hline
Base station antenna array height = 30 m\\ \hline
Mobile antenna height = 1.5 m\\ \hline
Precoding/Decoding: MR, ZF\\ \hline
Carrier spectral bandwidth = 50 MHz  \\ \hline
Carrier frequency = 60 GHz  \\ \hline
Cell radius = 200 m \\ \hline
Propagation model: line-of-sight\\ \hline
\end{tabular}
\end{table}

We note that at 60 GHz band, with half wavelength arc separation between antennas, the diameter of a 4096-antenna circular array is about $3.26$ meters. 

\begin{comment}
\subsection{Antenna Downtilt Model}
As in all practical deployments, antenna downtilts are necessary in multi-cell scenarios. We shall use the downtilt model proposed in \cite{G_2008}, where the vertical antenna pattern (i.e., gain in dB) is given by
$$
\Gamma (\theta) =\max\left\{ -12\left({\theta-\theta_{\sst{tilt}}\over \mbox{HPBW}}\right)^2, \mbox{SSL}\right\}
$$
where $\theta\in [-90^o, 90^o]$ is the negative elevation angle relative to the horizontal plane $\theta=0^o$, HPBW is the half-power beam width, SSL is the side lobe level. $\theta_{\sst{tilt}}$ is the downtilt angle.

\subsection{Comparison with Single Cell}
We compare the 95\% likely user throughput between multi-cell and single cell systems. Single cell with same number of service antennas serving same number of users, in the same area. 

In single cell setting, we see that ZF is power limited, with same radiated power, large bandwidth can results in smaller throughput. 
e.g., 50MHz bandwidth has higher throughput than 200MHz bandwidth, at R=100m.
\end{comment}

\section{Conclusions}

Under the assumption of LoS propagation, explicit formulas for the 
 downlink and uplink SINRs are provided for a multi-cell Massive MIMO system employing either MR or ZF linear processing. These
  formulas provide a means for readily analyzing system performance of Massive MIMO deployments in LoS scenarios. 

In a LoS propagation environment, inter-cell interference can be quite severe, and must be actively mitigated. Several approaches can be considered: 1) antenna downtilt, 2) using some degrees of freedom to null out interference, 3) since spectrum is more plentiful in the mmWave band, a frequency reuse factor greater than 1 may be considered.

As shown in \cite{YM:17:COM}, in a LoS propagation environment, the channel correlation between users can be quite high, which severely compromises the system throughput performance. Therefore, channel correlation must be minimized.  Channel correlation can be minimized by the following means: 1) schedule users with high channel correlation in different time slots, 2) 
if there are users with high channel correlation assigned to one base station, re-assign some of them to other  base stations.


%\bibliographystyle{IEEEtran}
%\bibliography{IEEEabrv,HienNgoBiblio}

\end{document}